\documentclass[proof]{WileyASNA-v1}

\usepackage[utf8]{inputenc}

\articletype{Article Type}%

\received{}
\revised{}
\accepted{}

\begin{document}

\title{Observability of ultraviolet N{\sc i} lines in the atmosphere of transiting Earth-like planets}

\author[1,2]{M. E. Young*}

\author[1]{L. Fossati}

\author[3]{C. Johnstone}

\author[4]{M. Salz}

\author[1]{H. Lichtenegger}

\author[5]{K. France}

\author[1]{H. Lammer}

\author[1]{P. E. Cubillos}

\authormark{YOUNG \textsc{et al}}

\address[1]{\orgdiv{Space Research Institute (IWF)}, \orgname{Austrian Academy of Science}, \orgaddress{\state{Styria}, \country{Austria}}}

\address[2]{\orgdiv{Department of Physics}, \orgname{University of Oxford}, \orgaddress{\state{South East England}, \country{UK}}}

\address[3]{\orgdiv{Department of Astrophysics}, \orgname{University of Vienna}, \orgaddress{\state{Vienna}, \country{Austria}}}

\address[4]{\orgdiv{Hamburg Observatory}, \orgname{University of Hamburg}, \orgaddress{\state{Hamburg}, \country{Germany}}}

\address[5]{\orgdiv{Laboratory for Atmospheric and Space Physics}, \orgname{University of Colorado}, \orgaddress{\state{Colorado}, \country{USA}}}

\corres{*M. E. Young. \email{mitchell.young@physics.ox.ac.uk}}

\presentaddress{Space Research Institute (IWF), Schmiedlstra{\ss}e 6, 8042 Graz, Austria}

\abstract{Nitrogen is a biosignature gas that cannot be maintained in its Earth-like ratio with CO$_2$ under abiotic conditions. It has also proven to be notoriously hard to detect at optical and infrared wavelengths. Fortunately, the ultraviolet region, which has only recently started being explored for terrestrial exoplanets, may provide new opportunities to characterise exoplanetary atmospheric nitrogen. In this work, the future prospects for detecting atomic nitrogen absorption lines in the transmission spectrum of an Earth-like planet orbiting in the habitable zone of a Sun-like star with LUVOIR are explored. Using the non-local thermodynamic equilibrium spectral synthesis code Cloudy, we produce a far-ultraviolet atomic transmission spectrum for an Earth-Sun-like system, and identify several nitrogen features, including both N{\sc i} and N{\sc ii} lines. We calculate the number of transits required for 1$\sigma$ and 3$\sigma$ detections of the planetary N{\sc i} $\lambda1200$ triplet signal with the G120M grating of the LUMOS spectrograph designed for LUVOIR, as a function of distance to the system and stellar ultraviolet emission. The minimum number of transit observations necessary for 1$\sigma$ and 3$\sigma$ detections of atomic N are 188 and 1685, respectively, for a system located at a distance of one pc with 100 times the Solar ultraviolet flux. Given that the orbital period of an Earth-Sun system is one year, it is not feasible to detect atomic N in the transmission spectrum for these systems. Future studies in this direction should therefore focus on Earth-like planets orbiting in the habitable zone of M dwarfs.}


\maketitle

\section{Introduction}\label{sec:intro}

The discovery and characterization of exoplanets that began in the 1990s \citep{wolszczan92, mayor95} launched the search for habitable worlds and extra-terrestrial life. Currently, only a hand-full of terrestrial planets have been found orbiting in the habitable zone (HZ) of solar-like stars \citep[e.g.][]{gillon17, dittmann17}, but planetary population statistics suggest that every star in the Galaxy should host at least one planet and that terrestrial planets are extremely common \citep[e.g.][]{cassan12, batalha13, dressing13, marcy14, bryson20}. The search for life and habitable worlds concentrates on the detection of biosignature gases found in the Earth's atmosphere, which have been extensively studied at optical and infrared (IR) wavelengths, both observationally and theoretically \citep[e.g.][]{ehrenreich06, stam08, palle09, kaltenegger09, vidal-madjar10, rauer11, garciamunoz12, sterzik12, betremieux13, misra14, arnold14, miles-paez14, yan15}.

Present-day Earth has a N-dominated atmosphere, characterised by large amounts of O and smaller amounts of H$_2$O and CO$_2$. \citet{snellen13} and \citet{arnold14} showed that H$_2$O and O$_2$/O$_3$ are directly detectable in the optical and IR with ground-based telescopes, particularly by targeting strong O$_2$ dimer features and O$_2$ bands. However, the sole detection of O in an exoplanetary transmission spectrum does not ensure the presence of life, as large amounts of oxygen can also be produced abiotically \citep[e.g.][]{schindler00S, selsis02, segura07, harman15, Schwieterman2016}. 

N$_2$ is the main component of Earth's atmosphere. Its partial surface pressure is maintained by bacteria which, under anaerobic conditions, return N$_2$ from the biosphere to the atmosphere, primarily through a process called denitrification \citep{cartigny13, wordsworth16}. Denitrification acts against atmospheric N$_2$ weathering, which would deplete Earth's atmosphere of its N$_2$ content within 100 Myr \citep{wordsworth16}. Meanwhile, the low levels of CO$_2$ in the Earth's atmosphere are maintained by weathering and plate tectonics, without which the Earth would have ended up with a CO$_2$-dominated atmosphere such as that of Venus or Mars \citep{lammer18}. Observed together, large amounts of N$_2$ and small amounts of CO$_2$ directly suggest the presence of an Earth-like habitat and bacterial life, as well as the presence of plate tectonics, which in turn suggests the presence of a large-scale planetary magnetic field, necessary to protect the atmosphere from erosion by the stellar wind \citep{lammer18}.

However, N and N$_{2}$ are extremely difficult to detect, with only a few rather weak atomic and molecular spectral features outside of the far-ultraviolet (FUV) bandpass. In the optical, N$_2$ produces a weak Rayleigh scattering slope, which has been shown to be very hard to detect \citep{arnold14}. At IR wavelengths, the dimer O$_2$·N$_2$ collision feature may be a candidate for the detection of N \citep{misra14}, but it is dependent on the abundance of atmospheric O. The biosignature gases NO$_2$ and N$_2$O present hardly any detectable features \citep{rauer11}. Even with the next generation of large space telescopes, it will be difficult to detect N-bearing molecules in the optical and IR \citep{kaltenegger09}. 

The UV is a wavelength region that has been largely unexplored for biosignatures in the transmission spectra of terrestrial exoplanets until recently \citep[e.g.][]{arney16,Hu13,krissansen-totton16,meadows17,olson18,schwieterman18}. Even so, N and N$_2$, both of which have many features in the FUV, are largely ignored in favour of other biosignatures, such as O$_2$ and O$_3$, with a few exceptions \citep{schwieterman15}. \citet{betremieux13} presented a theoretical study of the Earth's transmission spectrum at ultraviolet (UV) and optical wavelengths, but did not consider atomic features as they require modelling the upper atmospheric layers (thermosphere and exosphere), and thus, accounting for non-local thermodynamic equilibrium (NLTE). \citeauthor{betremieux13} demonstrate that, blueward of 2000 \AA, O$_2$ produces a signature 20 times stronger than any other optical/IR feature, and they advocate for further investigations of the Earth's UV transmission spectrum for biosignatures. \citet{youngblood20} recently presented the first UV observation of Earth as a transiting exoplanet, obtained with Hubble STIS during a Lunar eclipse, and found ozone signatures in the near-UV that were stronger than any found at optical or IR wavelengths. 

Recent and coming years have seen and will see the start of operations of major ground- and space-based facilities, such as TESS \citep{ricker2015} and PLATO \citep{rauer2014}, with the purpose of systematically searching for Earth-like planets orbiting late-type stars. Large space telescopes, such as HabEx \citep{habex2020} and LUVOIR \citep{luvoir2019} recently submitted to the NASA Decadal Survey, have also been conceived for the further detection and atmospheric characterisation of Earth-like planets orbiting nearby solar-like stars. The LUVOIR telescope, in particular, is foreseen to carry on-board the LUVOIR Ultraviolet Multi-Object Spectrograph \citep[LUMOS;][]{france2019}, which is perfectly poised to investigate planetary UV transmission spectra. 

In this work, we present the synthetic FUV transmission spectrum of an Earth-like planet based on models extending up to the outer layers of the planetary exosphere and accounting for NLTE effects. The advantage of UV atomic features over molecular features is that they form high in the planetary atmosphere and thus lead to large transit signals. However, this advantage comes at the expense of stellar signal-to-noise, because UV transit signals are seen against typically weak chromospheric and/or transition region stellar emission lines. With this in mind, we further investigate the future detectability of atomic N features in Earth-like atmospheres at UV wavelengths employing the LUMOS instrument and considering the LUVOIR-B (hereafter, called LUVOIR for simplicity) telescope configuration carrying an 8-m segmented primary mirror \citep{luvoir2019}.

\section{Synthetic Transmission Spectrum}\label{sec:trans_spec}

To synthesize the transmission spectrum of an Earth-like exoplanet, we follow the methodology outlined in \citet{young20}, using the NLTE spectral synthesis and plasma simulation code Cloudy \citep[v17.01;][]{ferland17}. Cloudy is primarily a hydrogen based code, and has not been extensively tested for environments heavily dominated by other elements such as an Earth-like N-dominated atmosphere. Therefore, in contrast to our previous work, rather than allow Cloudy to compute an atmospheric structure, we choose to employ a one dimensional (1-D) atmospheric structural model of Earth \citep{johnstone18} in our simulations. 

The 1-D structural model, extending from an altitude of 65 km at the lower boundary to an upper boundary at the exobase (475 km), includes number densities of seven elemental species (H, He, C, N, O, Ar and Cl) and 29 molecular species composed of these elemental constituents, as well as the temperature-altitude profile. To obtain the transmission spectrum, we map these 1-D model properties onto concentric spherical shells of radii equal to $1\,R_\oplus$ plus the model altitudes, and assume no day to night nor equator to pole variations. Using the quiet Sun irradiance spectrum \citep{woods09} as the illuminating source, Cloudy computes radiative transfer through the limb of the atmosphere at different altitudes. These altitude dependent limb transmission spectra are summed to produce the total transmission spectrum, weighted by the relative area of the stellar disc covered by the line-of-sight projection of the corresponding layer of the planetary atmosphere. The final transmission spectrum is output at a spectral resolving power of R\,=\,100\,000 over the 920$-$2700 \AA\ wavelength range.

\begin{figure}
\includegraphics[width=\columnwidth]{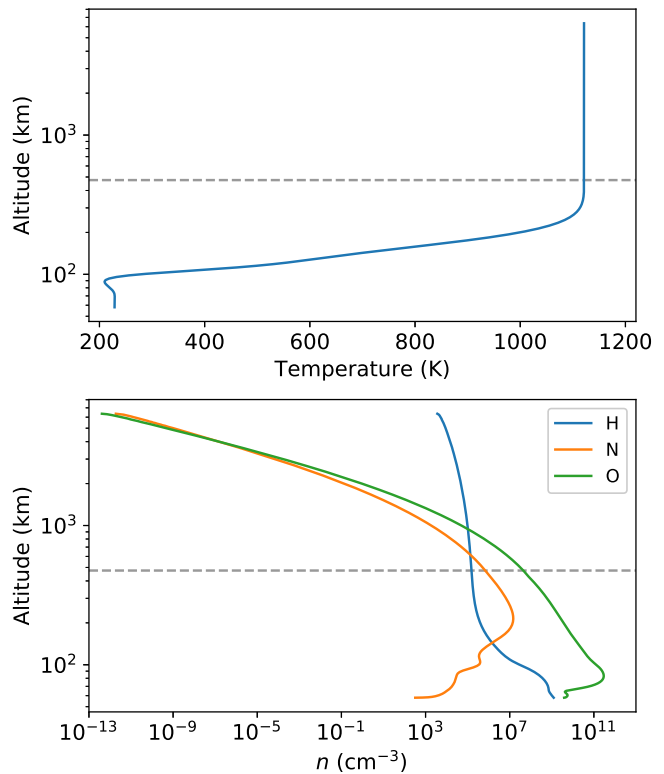}
\caption{1-D Earth-like atmospheric structural model employed in this work. $Top$: Temperature profile. $Bottom$: Elemental number density profiles, including atoms and ions, but not molecules. The grey dashed lines in each panel indicate the position of the exobase, hence the joint between the two models presented in the text. \label{fig:profiles}}
\end{figure}

Using this atmospheric structure in our spectral synthesis, we find that several spectral features in the transmission spectrum are still completely opaque at the altitude of the exobase. Therefore, we extend the temperature profile and density profiles of elemental H, N, and O to an altitude of $1\,R_\oplus$ above the planetary surface, following the Monte Carlo code described in \citet{pfleger2015} and \citet{vidotto2018}. To extend the profiles, particles are launched from the exobase with a Maxwell velocity distribution in isotropic upward directions and are assumed to move along collisionless Keplerian orbits above the exosphere. Any particles that pass the upper boundary are considered lost and removed from the simulation. The composite temperature profile and elemental H, N, and O density profiles are presented in Fig.~ \ref{fig:profiles}.

\begin{figure}
\includegraphics[width=\columnwidth]{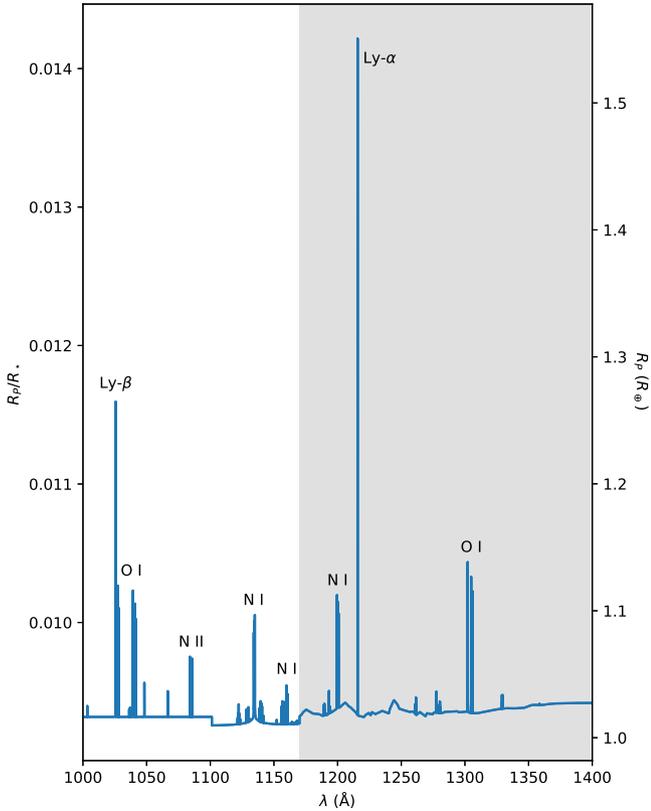}
\caption{Far-UV transmission spectrum of an Earth-like planet orbiting a Solar-like star. Prominent features include the N{\sc i} $\lambda1134$ and $\lambda1200$ triplets, N{\sc i} $\lambda1159.82$ and $\lambda1160.94$ lines, N{\sc ii} $\lambda1083.990$, $\lambda1084.580$, $\lambda1085.546$ and $\lambda1085.701$ lines, O{\sc i} $\lambda1305$, $\lambda1305$, and $\lambda1305$ triplets, Lyman series lines, and numerous weaker O lines. The grey shaded region indicates where we have added the \citet{betremieux13} molecular continuum. \label{fig:spectrum}}
\end{figure}

\citet{betremieux13} demonstrated that the transmission spectrum of an Earth-like atmosphere displays heavy absorption by molecular features at UV wavelengths, primarily O$_2$ and O$_3$, with some NO$_2$ and H$_2$O appearing as the wavelengths transition to the visible. Because we do not compute molecular absorption with Cloudy, we assume the molecular continuum of \citet{betremieux13}, and overlay it onto our atomic transmission spectrum. The composite transmission spectrum is presented in Fig.~\ref{fig:spectrum}.

In addition to Ly-$\alpha$ and Ly-$\beta$ (which extend to radii of 1.551 and 1.265 $R_\oplus$, respectively), three strong O{\sc i} triplets, and numerous weak O{\sc i} lines, several atomic N absorption features are identified in the transmission spectrum: two N{\sc i} triplets, the $\lambda1134$ triplet (1134.165 \AA, 1134.415 \AA, 1134.980 \AA) and the $\lambda1200$ triplet (1199.55 \AA, 1200.22 \AA, 1200.71 \AA); two individual N{\sc i} lines, $\lambda1159.82$ and $\lambda1160.94$; and 4 individual N{\sc ii} lines, $\lambda1083.990$, $\lambda1084.580$, $\lambda1085.546$ and $\lambda1085.701$. Of these features, only the N{\sc i} $\lambda1200$ triplet is covered by the range of the \citeauthor{betremieux13} molecular continuum, so it is this feature we focus our analysis on. Furthermore, among these features, the $\lambda$1200 triplet is the one for which the LUMOS effective area is largest.

\section{Synthetic Observations}\label{sec:obs}

To assess the detectability of the N{\sc i} $\lambda1200$ triplet with LUVOIR, we calculate the number of transits necessary to confirm 1$\sigma$ and 3$\sigma$ detections for systems with 1, 2, 5, 10, and 100 times the Solar far-UV (FUV) flux, set at distances of 1, 5, 10, and 25 pc from Earth. In all cases, we set the stellar and planetary radii equal to those of the Sun and Earth, respectively, and set the orbital semi-major axis to 1 AU. In doing this, we make five simplifying assumptions in our modelling and analysis.

\begin{enumerate}
    \item In the cases where the UV flux is greater than Solar, the planetary atmospheric structure and associated spectral absorption is unchanged.
    \item The transit impact parameter is 0, i.e. the planet transits through the center of the stellar disc, maximizing the available integration time during transit.
    \item The stellar limb effects have a negligible effect on the transmission spectrum or, at least, can be accounted for when generating the transmission spectrum.
    \item Variations in the stellar activity are negligible on the transit timescale.
    \item The system has a large enough radial velocity to distinguish between planetary and ISM absorption of N features.
\end{enumerate}

These assumptions are approximations only. Stars emitting large amounts of UV flux will be highly active, flaring significantly more often than the Sun. Likewise, it is highly unlikely that an Earth-like planet would maintain its atmosphere in a similar state to the model considered for an extended period of time under very enhanced UV flux conditions. It would be more heavily ionized, as well as hotter and more extended, which would imply a greater mass-loss rate. Stellar limb effects will have an impact on the transmission spectrum during the transit, but as no high resolution FUV limb spectra are available, and stellar atmospheric models do not properly treat the chromosphere and transition region, limb effects cannot be properly estimated here. Additionally, as the continuum would be limb darkened while the stellar emission lines would be limb brightened, the absorption signal would be strengthened while transiting the limb, rather than weakened.

For the hypothetical systems we are investigating, a systemic velocity in excess of $\pm50$ km/s relative to the ISM would be required to separate the planetary absorption signal. To assess the likelihood of finding such a system, we searched the GAIA DR2 catalog \citep{gaia16, gaia18} for main sequence late-type stars ($R_\star\leq1.5 R_\odot$, $T_{\rm eff}\leq6500$~K) with $|v_{\rm R}|\geq50$~km/s, within 25 pc. Setting these constraints, approximately 60 stars were identified as having the necessary systemic velocity. Assuming the conservative estimate of the occurrence rate of small planets around Sun-like stars from \citet{bryson20} ($\eta_\oplus = 0.37^{+0.48}_{-0.21}$), approximately 25 of these stars could host Earth-sized planets orbiting in the habitable zone, although these may not necessarily transit their hosts as seen from Earth.

For the LUMOS instrument, there are two gratings that include the N{\sc i} $\lambda1200$ triplet in their wavelength ranges, the medium spectral resolution G120M grating (R=40\,000) and the low resolution G155L (R=11\,550) grating \citep{france17,france2019}. We choose to simulate our transits using the G120M grating, which delivers the higher spectral resolution. We also note that both the G120M and G155L gratings cover the wavelengths of all the additional N{\sc i \& ii} spectral features identified in Sec.~\ref{sec:trans_spec}.

\begin{figure}
\includegraphics[width=\columnwidth]{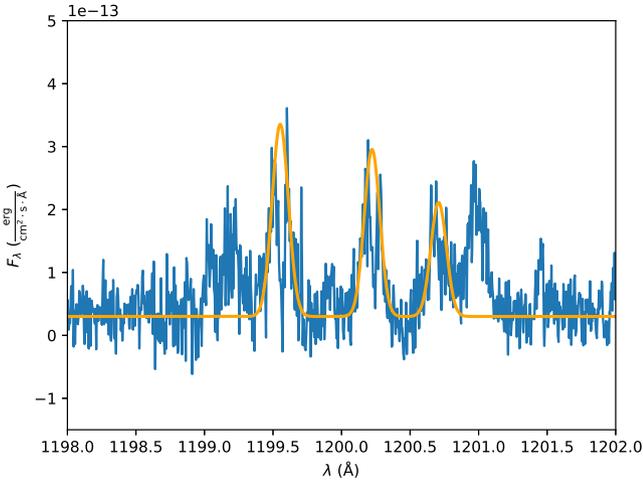}
\caption{Observed high resolution ($R$\,=\,52\,000) $\alpha$-Cen A N{\sc i} $\lambda1200$ triplet (blue), with Gaussian proxies (orange). The ISM absorption in the line cores is not considered when fitting the Gaussians. \label{fig:Gaussians}}
\end{figure}

To separate the stellar emission signal from the noise in the stellar spectrum, we fit the stellar N{\sc i} $\lambda1200$ triplet lines with three Gaussian profiles to produce noise free emission line proxies. The quiet Sun irradiance spectrum has a low spectral resolution of 1200 \AA, well below the resolution of the G120M grating, and the three lines are blended together into a single feature. To address this, for the spectrum of the background star we employ a high resolution ($R$\,=\,52\,000) spectrum of the solar analogue $\alpha$-Cen A obtained with the STIS spectrograph on-board HST and downloaded from the archive. Figure~\ref{fig:Gaussians} shows the observed high resolution $\alpha$-Cen A N{\sc i} $\lambda1200$ triplet and the Gaussian proxies fit. We note that the observed spectrum displays ISM absorption in the cores of the emission lines. To avoid having this absorption contaminating our signal-to-noise ratio (SNR) calculations, line cores are not considered when fitting the Gaussians to the observations, removing the additional absorption from our line proxies. By ignoring the ISM absorption in this fashion, this is necessarily an estimate applicable only to high radial velocity systems.

For microchannel plate detectors like those LUMOS uses in the far-UV (FUV), the SNR for bright targets can be approximated as the square root of the number of photons detected in a line. We convolve our line proxies to the resolution of the G120M grating and calculate the SNR for each line as
\begin{equation}
    SNR = \int_{\lambda_0-\frac{1}{2}FWHM}^{\lambda_0+\frac{1}{2}FWHM}\sqrt{F_\lambda \times \frac{d^2}{D^2} \times EA_\lambda \times T_{dur} \times \gamma}\,\,d\lambda\,,
\end{equation}
where $\lambda_0$ is the central wavelength of the given line, $FWHM$ is the full width of the line at half the maximum amplitude, $F_\lambda$ is the monochromatic flux spectrum in erg\,cm$^{-2}$\,s$^{-1}$\,\AA$^{-1}$, $d$ is the distance to $\alpha$\,Cen A (1.35 pc), we set $D$ as the distance to the system we are evaluating (in pc), $EA_\lambda$ is the effective area of the detector per wavelength (in $cm^2$), $T_\mathrm{dur}$ is the transit duration (in s), and $\gamma=5.03\times10^7\,\lambda\,F_\lambda$ is a conversion factor from energy flux to photon flux. 

To calculate the transit duration, we use
\begin{equation}
    T_\mathrm{dur}=\frac{P}{\pi}\sin^{-1}{\frac{\sqrt{(R_\star+R_{\rm p})^2-a^2\cos^2{i}}}{a}}\,,
\end{equation}
where $P\,=\,3.1556952\times10^7$ s is the orbital period (i.e., 1 year), $R_\star\,=\,R_\odot=6.975\times10^5$ km is the stellar radius, $R_{\rm p}\,=\,R_\oplus=6.3781\times10^3$ km is the planetary radius, $a\,=\,1\,AU=1.496\times10^8$ km is the orbital semi-major axis, and $i\,=\,90^o$ is the orbital inclination. Therefore, following our assumption of central transit,
\begin{equation}
    T_\mathrm{dur}=\frac{P}{\pi}\sin^{-1}{\frac{R_\star+R_{\rm p}}{a}}\,.
\end{equation}

To determine the planetary absorption signals of each N{\sc i} triplet, we convolved the square of the transmission spectrum (the transit depth) to the G120M grating resolution and integrated it over the same wavelength intervals as in the SNR integration, such that
\begin{equation}
{\rm signal} = \int_{\lambda_0-\frac{1}{2}FWHM}^{\lambda_0+\frac{1}{2}FWHM}\left(\frac{R_{\rm p}}{R_\star}\right)_\lambda^2 d\lambda\,,
\end{equation}
where $(\frac{R_{\rm p}}{R_\star})_\lambda$ is the synthetic planetary transmission spectrum. The total SNR and signal for a single transit is then the sum of the SNRs and signals for each of the three triplet lines. We note that for small terrestrial planets with realistic rotational velocities, the broadening of planetary absorption signals is minimal, and is not considered in this analysis.

For a 1$\sigma$ detection, we require a SNR ratio of $1/$signal, and likewise a SNR of $3/$signal for a 3$\sigma$ detection. Because SNR scales as the square root of the number of observations made, the number of transits needed to achieve a detection is then 
\begin{equation}
    N = \left(\frac{X/{\rm signal}_{t}}{{\rm SNR}_{t}}\right)^2\,,
\end{equation}
where X can either take the value of 1 or 3 for 1$\sigma$ or 3$\sigma$ detections, respectively, and ${\rm signal}_t$ and ${\rm SNR}_t$ are the signal and SNR for a single transit, respectively. The numbers of transits necessary to achieve detections of $N$ in the transmission spectrum are presented in Table~\ref{tab:transits}.

\begin{table*}
\centering
\caption{Number of transits necessary for a 1$\sigma$ (3$\sigma$) detection of atomic nitrogen absorption in the atmosphere of an Earth-like exoplanet orbiting a Solar-like star as a function of stellar FUV emission flux and distance to the star.}
\label{tab:transits}
\begin{tabular}{c|ccccc}
\hline
\noalign{\smallskip}
Distance to & \multicolumn5c{UV Flux relative to Solar}\\
System (pc) & 1x & 2x & 5x & 10x & 100x\\
\noalign{\smallskip}
\hline
\noalign{\smallskip}
1 & 1.9$\times10^4$ (1.7$\times10^5$) & 9357 (8.4$\times10^4$) & 3743 (3.4$\times10^4$) & 1872 (1.7$\times10^4$) & 188 (1685) \\
5 & 4.7$\times10^5$ (4.2$\times10^6$) & 2.3$\times10^5$ (2.1$\times10^6$) & 9.4$\times10^4$ (8.4$\times10^5$) & 4.7$\times10^4$ (4.2$\times10^5$) & 4679 (4.2$\times10^4$) \\
10 & 1.9$\times10^6$ (1.7$\times10^7$) & 9.4$\times10^5$ (8.4$\times10^6$) & 3.7$\times10^5$ (3.4$\times10^6$) & 1.9$\times10^5$ (1.7$\times10^6$) & 1.9$\times10^4$ (1.7$\times10^5$) \\
25 & 1.2$\times10^7$ (1.1$\times10^8$) & 5.8$\times10^6$ (5.3$\times10^7$) & 2.3$\times10^6$ (2.1$\times10^7$) & 1.2$\times10^6$ (1.1$\times10^7$) & 1.2$\times10^5$ (1.1$\times10^6$) \\
\noalign{\smallskip}
\hline
\end{tabular}
\end{table*}

\section{Discussion and conclusion}

Starting from a 1D atmospheric structure model of the Earth, we have computed an UV atomic transmission spectrum using Cloudy. In this transmission spectrum, we have identified the presence of eight N{\sc i} lines (two triplets and two individuals) and four N{\sc ii} lines, of which we are certain the N{\sc i} $\lambda1200$ triplet is optically thick above the molecular continuum. To assess the detectability of atomic N in a transmission spectrum like this one, we simulated LUVOIR LUMOS observations for hypothetical planetary systems at several distances from Earth, with different levels of stellar UV flux, assuming a constant atmospheric structure, an impact parameter of zero, and negligible stellar limb effects and activity. We find that, for an Earth-like planet in orbit around a Solar-like star, the detection of atomic N is not feasible even with a next generation facility such as LUVOIR. Even in the ideal scenario, a 1$\sigma$ detection requires 188 transits be observed, and a 3$\sigma$ detection under the same conditions requires 1685 transits be observed.

Because we placed the planets in Earth-like orbits, the orbital periods are equal to one year, meaning the number of transits required for the detections is also the number of years over which the observations would need to be made. Even in the ideal scenario of a system located at a distance of 1 pc, with a stellar UV flux equal to 100 times solar, a 1$\sigma$ detection would take 188 years, longer than a human lifetime and certainly longer than the LUVOIR mission lifetime.

On one hand, some of the assumptions listed in Sec.~\ref{sec:obs} imply that the values listed in Table~\ref{tab:transits} should be considered lower limits on the number of transits required for detecting atomic N. For example, we assumed an impact parameter equal to zero, but it is more likely that the parameter would be larger than this, leading to shorter available integration times. Also, ISM absorption of stellar emission lines would reduce the SNR of an observation, increasing the number of transits required to detect N{\sc i}, as not many systems hosting Earth-sized planets have the radial velocity required to shift the planetary absorption away from the ISM absorption. A stellar UV flux higher than Solar would probably increase the size of the occluding planetary atmosphere, but also increase ionisation, creating competing effects on the detectability of atomic N. On the other hand, the lack of molecular continuum below 1170\,\AA\ prevents us from including the other available atomic nitrogen lines when computing the detectability. As a matter of fact, in addition to the N{\sc i} $\lambda1200$ triplet, there are a total of five other N{\sc i} lines and four N{\sc ii} lines in the operating range of LUMOS, with EAs at least 45\% the size of the EA at 1200~\AA. Conservatively, if these additional lines increase the signal by at least a factor of two, that reduces the number of transits required for detection by a factor of four. 

While these factors likely do not grant significant improvement for the detection of atomic N in the situation of an Earth-like planet around a Solar-like star, Earth-like planets around smaller stars such as M dwarfs may provide a better opportunity. M dwarf stars typically have radii between $\sim0.6~R_\odot$ for early M stars and $\sim0.08~R_\odot$ for late M stars, with bolometric luminosities $<10\%~L_\odot$ \citep{kaltenegger09}, but can maintain UV fluxes in excess of the Solar UV flux, by more than two orders of magnitude in some cases, and may indeed still be able to support life under these conditions \citep{omalley-james19}. 

Take, for example, a star like the M1.5 V star GJ667\,C, a typical early M dwarf, which displays a stellar UV emission in the N{\sc i} $\lambda1200$ triplet and in Ly-$\alpha$ similarly to $\alpha$ Cen A \citep{france16}. For an Earth-like planet orbiting in the HZ of GJ667\,C (0.172 AU), the orbital period would be $\sim1.5$ months, equivalent to eight transits a year. Additionally, GJ667\,C has a stellar radius of 0.42~$R_\odot$, meaning an Earth-like planet would block a proportionately larger fraction of the stellar disc, increasing the strength of the absorption signal. Admittedly, the transit duration would be shorter, $\sim3$ hours compared to the $\sim13$ hour duration for a Sun-like star, but overall, a N detection could be made in a shorter time than for a Sun-like star.

Recently, the discovery of HD 21749c, the first Earth-sized planet discovered with TESS, was announced \citep{dragomir19}. While HD 21749c doesn't orbit in the HZ of its host star, a K4.5 V star at 16 pc from Earth, the detection of this planet is a clear indication that TESS is able to find Earth-sized planets orbiting smaller stars. As the TESS mission continues, discoveries of additional Earth-sized planets will occur, potentially orbiting in the HZs of their host stars. These will be the targets on which we will have the first opportunities to attempt characterising the atmospheres of other planets with masses and sizes similar to those of the Earth. Our results clearly indicate that the search for nitrogen-dominated atmospheres outside of the solar system, at least for Earth-like planets, will have to focus on the optical and infrared bands. At these wavelengths, features of nitrogen-bearing molecules are scarce and small, but hopefully the amount of stellar photons will be large enough to overcome these limitations.

\section*{Acknowledgments}

The authors would like to thank T. T. Koskinen for added insight on aspects of this work. This research has made use of NASA's Astrophysics Data System Bibliographic Services, and the NIST Atomic Spectra Database funded [in part] by NIST's Standard Reference Data Program (SRDP) and by NIST's Systems Integration for Manufacturing Applications (SIMA) Program. The following software and packages were used in this work: \texttt{Cloudy v17.01} (\citealt{ferland17}); \texttt{Python v2.7}; \texttt{Python} packages \texttt{Astropy} \citep{astropy13}, \texttt{NumPy} \citep{numpy06,numpy11}, and \texttt{Matplotlib} \citep{matplotlib07}.

This work was funded by the \fundingAgency{Austrian Academy of Sciences (OeAW)} grant \fundingNumber{OeAW-Innovationsfonds IF201703.} MEY acknowledges funding from the \fundingAgency{European Research Council (ERC)} under the European Union’s Horizon 2020 research and innovation program under grant agreement \fundingNumber{No 805445.} We thank the anonymous referee for helpful comments that led to improve the manuscript.

\subsection*{Conflict of interest}

The authors declare no potential conflict of interests.

\bibliography{ms}

\end{document}